# Acquisition of high-quality three-dimensional electron diffuse scattering data

**Romy Poppe[a] and Joke Hadermann[a]**

[a]University of Antwerp, Department of Physics, Groenenborgerlaan 171, B-2020 Antwerp, Belgium

**Correspondence email**: romy.poppe@uantwerpen.be

**Funding information**: This work was supported by the Research Foundation Flanders (FWO Vlaanderen) [grant numbers G035619N and G040116N].

**Abstract**: The diffraction patterns of crystalline materials with local order contain sharp Bragg reflections as well as highly structured diffuse scattering. The instrumental requirements, experimental parameters and data processing techniques for obtaining high-quality diffuse scattering data have previously been determined for X-ray and neutron diffraction, but not yet for electron diffraction. In this study, we show that the spatial resolution of the diffuse scattering in three-dimensional electron diffraction (3D ED) data depends on various effects, including the convergence of the electron beam, the point spread function of the detector and the crystal mosaicity. In contrast to single-crystal X-ray diffraction, the detector point spread function for 3D ED is broader for a hybrid pixel detector than for a CCD. In our study, we also compare the diffuse scattering in 3D ED data with the diffuse scattering in single-crystal X-ray diffraction data and show that the diffuse scattering in 3D ED data can be obtained with a quality comparable to that from single-crystal X-ray diffraction. As electron diffraction requires much smaller crystal sizes than X-ray diffraction, this opens up the possibility to investigate the local structure of many technologically relevant materials for which no crystals large enough for single-crystal X-ray diffraction are available.

## 1. Introduction

In conventional crystallography, it is assumed that a crystalline material consists of a three-dimensional array of identical units. Real materials, however, only approximate this ideal and their diffraction patterns contain, in addition to sharp Bragg reflections, a weak continuous background known as diffuse scattering (Welberry & Weber, 2016). Diffuse scattering occurs whenever there are deviations from the average structure – that is, when the crystal is disordered. When the deviations from the average structure are ordered on a local scale, they give rise to highly structured diffuse scattering. Examples of parameters that can be refined from the intensity distribution of the diffuse scattering are the number of stacking faults, correlations between neighbouring atoms and local atomic displacements. Because the properties of many materials do not only depend on the average structure, but also on the disorder, quantitative analysis of diffuse scattering is essential for understanding and optimizing material properties.

Three-dimensional electron diffraction (3D ED) was developed in 2007 (Kolb *et al.*, 2007, 2008). The main advantage of 3D ED is that it allows to determine the crystal structure of materials for which no crystals large enough for single-crystal X-ray diffraction are available (Palatinus *et al.*, 2017). The diffuse scattering in electron diffraction data has been analysed both qualitatively (Withers *et al.*, 2003, 2004; Fujii *et al.*, 2007; Goodwin *et al.*, 2007; Brázda *et al.*, 2016; Zhao *et al.*, 2017; Neagu & Tai, 2017; Gorelik *et al.*, 2023) and quantitatively (Krysiak *et al.*, 2018, 2020; Poppe *et al.*, 2022). For the quantitative analysis of diffuse scattering, however, high-quality diffuse scattering data are needed. The instrumental requirements, experimental parameters and data processing techniques for obtaining high-quality diffuse scattering data have previously been determined for X-ray and neutron diffraction (Welberry & Weber, 2016), but not yet for electron diffraction, and is the scope of this study.

## 2. Experimental section

### 2.1. Synthesis

The samples that were used in this study were previously used by (Roth *et al.*, 2021) and are referred to as the 'SC-0.81' and 'Q-0.84 #2' samples. Two different synthesis methods were used to prepare these samples. The 'SC-0.81' sample has nominal stoichiometry $Nb_{0.81}CoSb$ and was slowly cooled (SC) using an induction furnace. The 'Q-0.84 #2' sample has nominal stoichiometry $Nb_{0.84}CoSb$ and was thermally quenched (Q) from the melt. Details of the synthesis can be found in (Roth *et al.*, 2021) for the slowly cooled sample and in (Yu *et al.*, 2018) for the thermally quenched sample. The thermally quenched sample $Nb_{0.84}CoSb$ (Q-0.84 #2) only has short-range Nb-vacancy order, whereas the slowly cooled sample $Nb_{0.81}CoSb$ (SC-0.81) also has long-range Nb-vacancy order.

### 2.2. Electron diffraction data collection

Samples for electron diffraction data collection were prepared by dispersing the powder in ethanol. A few droplets of the suspension were deposited on a copper grid covered with an amorphous carbon film. Ultra-thin amorphous carbon grids were used to reduce the experimental background.

In-zone selected area electron diffraction (SAED) patterns were acquired with an FEI Tecnai G2 electron microscope operated at 200 kV using an FEI Eagle 2k CCD camera (2048 x 2048 pixels with 16-bit dynamic range). In-zone precession electron diffraction (PED) patterns were acquired with a precession angle of 1° using a DigiSTAR precession device from NanoMEGAS.

Three-dimensional electron diffraction (3D ED) data were acquired with an aberration-corrected cubed FEI Titan 80-300 electron microscope operated at 300 kV using a GATAN US1000XP CCD camera (4096 x 4096 pixels with 16-bit dynamic range). One 3D ED data set was acquired using a Quantum Detectors MerlinEM hybrid pixel detector (512 x 512 pixels with 24-bit dynamic range). The crystal was illuminated in SAED mode with an exposure time of 1 s per frame. Electron diffraction patterns were acquired with a Fischione tomography holder (tilt range of ±80°), in a stepwise manner, using an in-house developed script. The 3D ED data were collected with a step size of 0.1° on crystals with a size of 200-3000 nm. For the larger crystals, only a thin part of the crystal was illuminated, which was recentred inside the aperture every few degrees. Energy filtered 3D ED data were acquired with a Quantum 966 Gatan Image Filter, with a slit width of 10 eV.

*PETS2* (Palatinus *et al.*, 2019) was used to process the 3D ED data including background subtraction of the individual frames, and applying symmetry with Laue class $m\bar{3}m$ in the reconstruction of the three-dimensional reciprocal lattice. The three-dimensional reciprocal lattice of all 3D ED data was indexed with a cubic unit cell with cell parameter a = 5.89864(3) Å and space group $F\bar{4}3m$ (Zeier *et al.*, 2017).

Details on the acquisition of the single-crystal X-ray diffraction data can be found in (Roth et al., 2021).

## 3. Results and discussion

### 3.1. Three-dimensional electron diffraction and in-zone electron diffraction

The main advantage of three-dimensional electron diffraction (3D ED) is that it allows the acquisition of three-dimensional electron diffuse scattering data with less multiple scattering compared to in-zone selected area electron diffraction (SAED) patterns. Another method that is often used to record the intensities of the Bragg reflections with reduced multiple scattering (Vincent & Midgley, 1994) is precession electron diffraction (PED). In PED, the electron beam is tilted away from the optical axis of the electron microscope by a certain angle (the precession angle, typically 1-3°) and rotated on the surface of a cone with the vertex fixed on the sample plane. The resulting PED pattern is obtained by integrating the intensities in the acquired off-axis electron diffraction patterns. When the crystal is oriented along a zone-axis, and the electron beam is tilted away from this zone-axis, the total number of possible paths is reduced, and consequently also the amount of multiple scattering.

Fig. 1 shows the *h0l* and *hhl* planes reconstructed from 3D ED data, in-zone SAED patterns, and in-zone PED patterns acquired on the same crystal. The diffuse circles in the *h0l* plane reconstructed from 3D ED show clear intensity modulations. By contrast, the diffuse circles in the in-zone SAED pattern have almost the same intensity everywhere due to multiple scattering. The intensity distribution of the diffuse scattering in in-zone PED patterns is very similar to the one in in-zone SAED patterns. The main difference is that the higher-order Bragg reflections have higher intensities in the in-zone PED patterns than in the in-zone SAED patterns. Bragg reflections are distinct points, whereas diffuse scattering is continuously distributed in reciprocal space. An in-zone PED pattern is obtained by integrating the intensities within a volume in reciprocal space (determined by the precession angle). Higher precession angles will thus decrease the spatial resolution of the observed diffuse scattering.

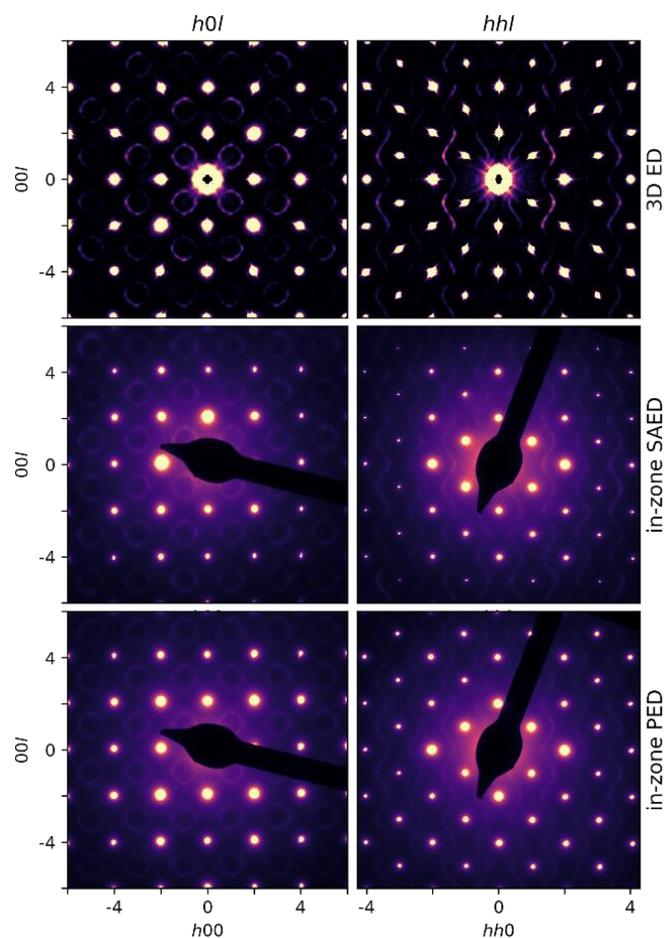

**Figure 1** Comparison of the *h0l* and *hhl* planes reconstructed from three-dimensional electron diffraction (3D ED) data, in-zone selected area electron diffraction (SAED) patterns, and in-zone precession electron diffraction (PED) patterns acquired on the same crystal. Data acquired on the thermally quenched sample (Q-0.84 #2).

### 3.2. Background subtraction

Because the Bragg reflections are many orders of magnitude stronger than the diffuse scattering, the acquisition of high-quality diffuse scattering data requires careful subtraction of the experimental background. Obtaining high-quality diffuse scattering data with negligible background contributions comprises three steps: (i) increasing intensities, (ii) reducing the experimental background and (iii) post-experimental elimination of the background (Welberry & Weber, 2016). The background of the in-zone SAED patterns and in-zone PED patterns in Fig. 1 has not been subtracted. For the 3D ED data, the background of the individual frames has been subtracted in *PETS2* before the reconstruction of the three-dimensional reciprocal lattice.

The background in electron diffraction data is due to inelastic scattering of the incoming electrons (thermal diffuse scattering), electrons scattered by the amorphous carbon film, and sensor intrinsic background noise of the CCD. Thermal diffuse scattering can be subtracted using an energy filter. Fig. 2 shows the *hhl* plane reconstructed from 3D ED data acquired with and without energy filter on the same crystal. The energy filter blocks inelastically scattered electrons with an energy loss of more than 10 eV. Except for using an energy filter, the experimental settings were identical for both 3D ED data sets. The *hhl* plane reconstructed from 3D ED data acquired without energy filter shows strong diffuse intensity bands. The diffuse intensity bands are weaker in the *hhl* plane reconstructed from 3D ED data acquired with energy filter but can still be observed. An energy filter with a slit width of 10 eV reduces the thermal diffuse scattering but does not entirely remove it. The *hhl* planes after background subtraction in *PETS2* are also shown in Fig. 2. After background subtraction in *PETS2*, the *hhl* plane looks similar for the energy-filtered and non-energy filtered data. The experimental background and the thermal diffuse scattering are subtracted, while the elastic diffuse scattering remains.

The background in *PETS2* is subtracted by using a median filter algorithm (Palatinus *et al.*, 2019). To ensure that the elastic diffuse scattering is not subtracted, the reflection size in *PETS2* should be chosen equal to or a bit larger than the width of the diffuse scattering contours. Another method to estimate the background, which is sometimes used for single-crystal X-ray diffuse scattering data, is to repeat the data acquisition under the same conditions but without the crystal illuminated by the beam. In the case of one-dimensional or two-dimensional diffuse scattering, the background intensity can also be estimated from the intensities of the surrounding voxels (Welberry & Weber, 2016). However, both methods do not subtract the thermal diffuse scattering, and would thus require energy-filtered 3D ED data.

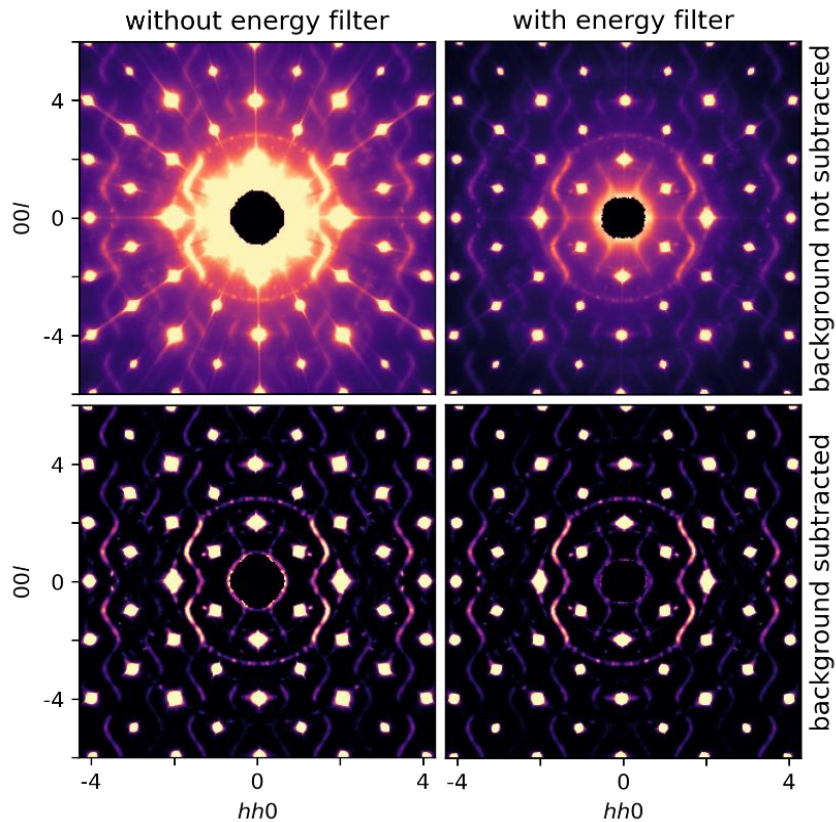

**Figure 2** Comparison of the *hhl* plane reconstructed from three-dimensional electron diffraction (3D ED) data acquired with and without energy filter, before and after background subtraction in *PETS2*. The circle passing through the (202) Bragg reflection is due to neighbouring crystals. Due to symmetry averaging with Laue class $m\bar{3}m$, each additional reflection in the *hhl* plane will appear four times. Data acquired on the thermally quenched sample (Q-0.84 #2).

### 3.3. Symmetry averaging

Fig. 3 shows the *h0l* and *hhl* planes reconstructed from 3D ED data before and after applying symmetry averaging with Laue class $m\bar{3}m$. The *h0l* and *hhl* planes before applying symmetry averaging have a missing wedge due to the limited tilt range (90° for the thermally quenched sample (Q-0.84 #2) and 100° for the slowly cooled sample (SC-0.81)). Applying symmetry averaging with Laue class $m\bar{3}m$ allows to fill the missing wedge in the three-dimensional reciprocal lattice. This is required for the calculation of the three-dimensional difference pair distribution function (3D-ΔPDF), which is a method that is often used to determine the origin of the diffuse scattering (Schaub *et al.*, 2007).

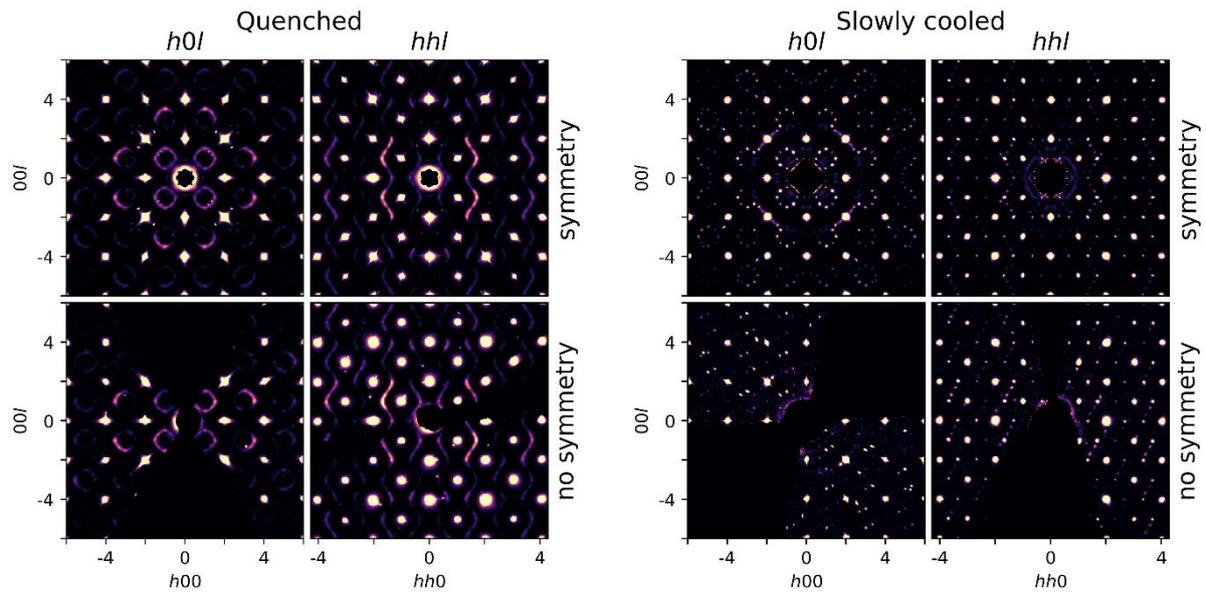

**Figure 3** Comparison of the *h0l* and *hhl* planes reconstructed from three-dimensional electron diffraction (3D ED) data before and after applying symmetry averaging with Laue class $m\bar{3}m$, both for the thermally quenched sample (Q-0.84 #2) and the slowly cooled sample (SC-0.81). The additional reflections between the Bragg reflections are due to neighbouring crystals. Due to symmetry averaging with Laue class $m\bar{3}m$, each additional reflection in the *h0l* plane will appear eight times, while each additional reflection in the *hhl* plane will appear four times.

The main advantage of 3D ED is that it allows to determine the crystal structure of materials for which no crystals large enough for single-crystal X-ray diffraction are available. The 3D ED data for the slowly cooled sample (SC-0.81) were acquired on a 150 nm sized crystal. Crystals with long-range Nb-vacancy order consist of twins with different orientations (Xia *et al.*, 2019). Each twin orientation gives rise to one pair of satellite reflections. Because some twin orientations are missing, not all satellite reflections are visible in the *h0l* and *hhl* planes before symmetry averaging. Symmetry averaging with Laue class $m\bar{3}m$ will thus introduce additional satellite reflections corresponding to the other twin orientations.

### 3.4. Convergence of the electron beam

In 3D ED, the crystal can be illuminated either in SAED mode or in nano electron diffraction (NED) mode (Gemmi *et al.*, 2019). In SAED mode, the incident electron beam is parallel, and the sample area used for collecting diffraction data is determined by the selected area aperture. In NED mode, a small C2 condenser aperture is inserted, and the sample area used for collecting diffraction data is determined by the beam size. The incident electron beam in NED mode is usually slightly convergent. Acquiring 3D ED data in SAED mode will thus improve the spatial resolution of the observed diffuse scattering.

Fig. 4 shows the *h0l* and *hhl* planes reconstructed from 3D ED data acquired on the same crystal in SAED mode and in NED mode. Due to the slightly convergent electron beam in NED mode, the higher-order Bragg reflections have higher intensities for the 3D ED data acquired in NED mode than for the 3D ED data acquired in SAED mode. For high-resolution experiments, 3D ED data should thus be acquired in SAED mode.

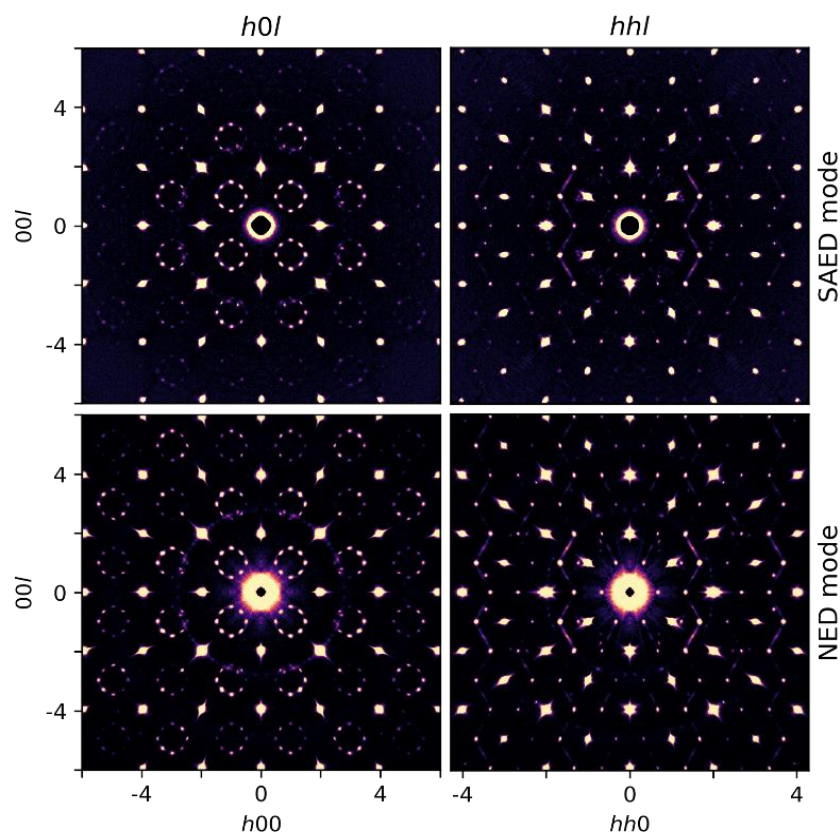

**Figure 4** Comparison of the *h0l* and *hhl* planes reconstructed from three-dimensional electron diffraction (3D ED) data acquired in selected area electron diffraction (SAED) and nano electron diffraction (NED) mode. Data acquired on the slowly cooled sample (SC-0.81).

### 3.5. Detector point spread function

Important detector performance characteristics for diffuse scattering measurements include a narrow detector point spread function, low sensor intrinsic background noise, and a high dynamic range (Welberry & Weber, 2016). Hybrid pixel detectors such as Pilatus (Kraft *et al.*, 2009), XPAD (Pangaud *et al.*, 2008) or Medipix (Gimenez *et al.*, 2011) are used for the acquisition of high-quality single-crystal X-ray diffuse scattering data at synchrotron sources (Welberry & Weber, 2016). The point spread function is essentially one pixel broad, they can be operated under zero intrinsic noise conditions, and the dynamic range is higher than for CCDs (Welberry & Goossens, 2016).

Fig. 5 shows the *hhl* plane reconstructed from 3D ED data acquired with a GATAN US1000XP CCD camera and a Quantum Detectors MerlinEM hybrid pixel detector on the same crystal. In contrast to single-crystal X-ray diffraction, the detector point spread function for 3D ED is broader for the hybrid pixel detector than for the CCD, which can be explained by charge sharing (Jakůbek, 2009). Electrons that fall in on the hybrid pixel detector create a charge cloud in the silicon sensor. Electrostatic repulsion and charge diffusion cause the charge cloud to expand, which explains why electrons are also detected by neighbouring pixels. The spatial resolution of the observed diffuse scattering is higher for 3D ED data acquired with a CCD than for 3D ED data acquired with a hybrid pixel detector, and for high-resolution experiments, 3D ED data should thus be acquired using a CCD.

The spatial resolution of diffuse scattering data acquired with a CCD detector depends on the thickness of the phosphor layer (Welberry & Weber, 2016). The disadvantage of using a CCD detector is that the dynamic range is lower than for hybrid pixel detectors. Consequenlty, the Bragg reflections close to the central beam were overexposed, giving rise to blooming artefacts. Besides, diffuse scattering data collected with a CCD contain sensor intrinsic background noise (Welberry & Weber, 2016). The development of hybrid pixel detectors optimized for electron diffraction measurements is thus necessary to improve the quality of the observed diffuse scattering in 3D ED data.

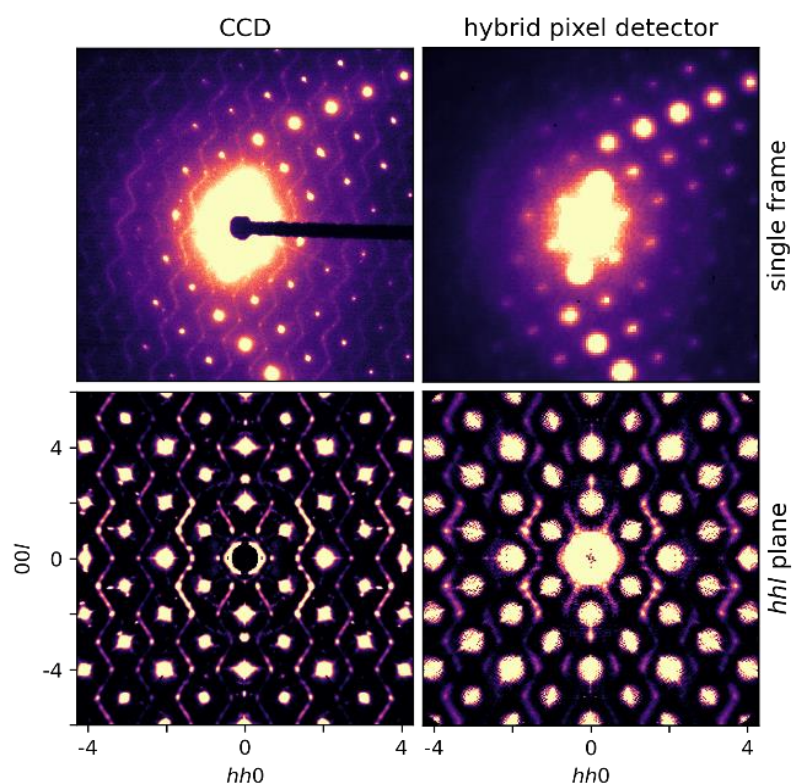

**Figure 5** Comparison of the *hhl* plane reconstructed from three-dimensional electron diffraction (3D ED) data acquired with a GATAN US1000XP CCD camera and a Quantum Detectors MerlinEM hybrid pixel detector on the same crystal. Data acquired on the thermally quenched sample (Q-0.84 #2).

### 3.6. Crystal mosaicity

Fig. 6 shows the *h0l* and *hhl* planes reconstructed from 3D ED data acquired on five different crystals of the thermally quenched sample (Q-0.84 #2). The angular broadening of the Bragg reflections is larger for crystal 4 than for crystal 5, which is due to differences in the crystal mosaicity. Crystals consist of domains in which the lattice planes are slightly misaligned. The larger the spread of lattice plane orientations, the larger the mosaicity. The spatial resolution of the observed diffuse scattering will be higher for crystals with a lower mosaicity.

Not all crystals of the thermally quenched sample (Q-0.84 #2) have identical diffuse scattering. All crystals have satellite reflections on top of the diffuse scattering, but their sharpness is different. The sharpness of the satellite reflections is related to the correlation length of the local Nb-vacancy order.

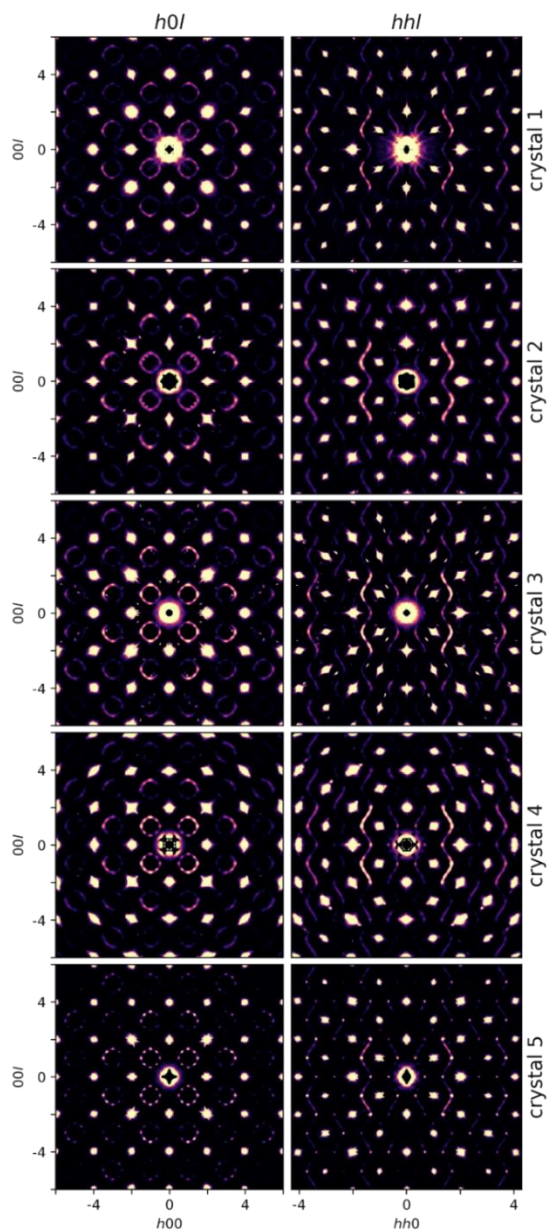

**Figure 6** Comparison of the *h0l* and *hhl* planes reconstructed from three-dimensional electron diffraction (3D ED) data acquired on five different crystals of the thermally quenched sample (Q-0.84 #2). The additional reflections between the Bragg reflections for crystals 2 and 3 are due to neighbouring crystals. Due to symmetry averaging with Laue class $m\bar{3}m$, each additional reflection in the *h0l* plane will appear eight times, while each additional reflection in the *hhl* plane will appear four times.

**3.7. X-ray and electron diffraction**

Fig. 7 shows the *hhl* plane and the planes 1, 2, 3 and 4 voxels above the *hhl* plane, reconstructed from single-crystal X-ray and 3D ED data acquired on the slowly cooled sample (SC-0.81). Each plane has a thickness of one voxel. The *hhl* plane reconstructed from 3D ED data shows additional satellite reflections compared with the *hhl* plane reconstructed from single-crystal X-ray diffraction data (satellite reflections indicated by the white circles). These additional satellite reflections have their maximum intensity in the plane four voxels above the *hhl* plane. The Bragg reflections are broader in the 3D ED data than in the single-crystal X-ray diffraction data, which was also observed in (Schmidt *et al.*, 2023). The spatial resolution of the observed diffuse scattering is thus lower for 3D ED than for single-crystal X-ray diffraction. Consequently, the reflections indicated by the white circles in the *hhl* plane reconstructed from 3D ED are from slightly above and below the *hhl* plane.

The spatial resolution of the diffuse scattering is determined by various effects, including the convergence of the beam, the point spread function of the detector and the crystal mosaicity. The main reason for the difference in spatial resolution between 3D ED and single-crystal X-ray diffraction is probably the use of a different type of detector. Other parameters that may influence the spatial resolution of the observed diffuse scattering are the monochromaticity of the beam, vibrations of the crystal or the instrument, the detector distance and the data collection step width (Boysen & Adlhart, 1987). For high-resolution experiments, 3D ED data should be collected with a step size of 0.1° or smaller (Welberry & Weber, 2016), which was the case in this study. To a first approximation, most of these effects broaden Bragg reflections isotropically and uniformly and the resolution function can be approximated by a Gaussian function (Weber & Simonov, 2012). For the refinement of local order parameters in *Yell* (Simonov *et al.*, 2014) and *DISCUS* (Proffen & Neder, 1997), resolution effects can be considered by convoluting the intensity of each voxel in the calculated data with a Gaussian function. The standard deviation of this Gaussian function can be estimated from the intensity profile of unsaturated Bragg reflections. However, angular broadenings caused by e.g., crystal mosaicity and radial broadenings caused by e.g., the spectral width of the beam cannot easily be corrected (Weber & Simonov, 2012) so the experimental minimization of resolution effects is essential.

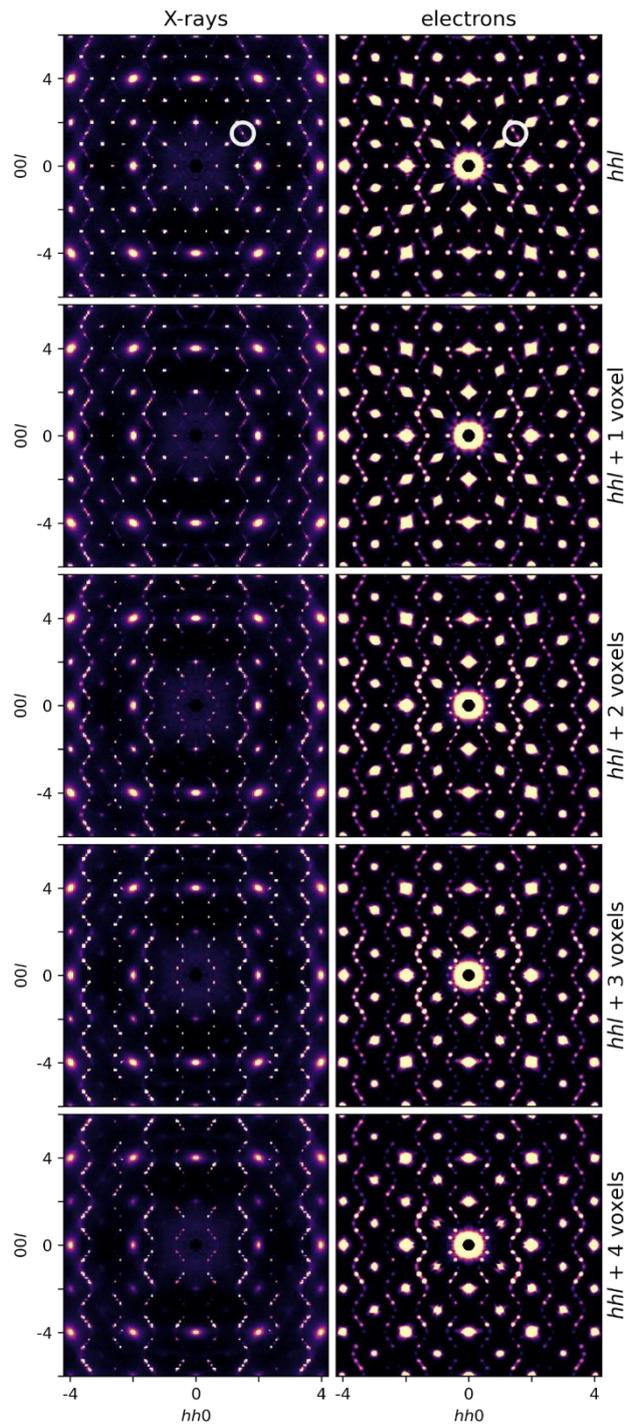

**Figure 7** Comparison of the *hhl* plane and the planes 1, 2, 3 and 4 voxels above the *hhl* plane, reconstructed from single-crystal X-ray diffraction and three-dimensional electron diffraction (3D ED) data. The *hhl* plane reconstructed from 3D ED data shows additional satellite reflections compared with the *hhl* plane reconstructed from single-crystal X-ray diffraction data (reflections indicated by the white circles). These additional satellite reflections have their maximum intensity four voxels above the *hhl* plane. Data acquired on the slowly cooled sample (SC-0.81).

## 4. Conclusion

Three-dimensional electron diffraction (3D ED) allows to acquire the three-dimensional diffuse scattering with less multiple scattering compared to in-zone electron diffraction patterns. The instrumental requirements, experimental parameters and data processing techniques for obtaining high-quality diffuse scattering data were investigated. We showed that the spatial resolution of the diffuse scattering in 3D ED data depends on various effects, including the convergence of the electron beam, the point spread function of the detector and the crystal mosaicity. In contrast to single-crystal X-ray diffraction, the detector point spread function for 3D ED is broader for a hybrid pixel detector than for a CCD.

The three-dimensional diffuse scattering in 3D ED data can be obtained with a quality comparable to that from single-crystal X-ray diffraction. However, the spatial resolution of the diffuse scattering in 3D ED data acquired with a CCD in a transmission electron microscope is still lower compared to the spatial resolution of the diffuse scattering in single-crystal X-ray diffraction data acquired with a hybrid pixel detector at a synchrotron source. The development of hybrid pixel detectors optimized for electron diffraction measurements is thus necessary to improve the spatial resolution of the observed diffuse scattering in 3D ED data.

As electron diffraction requires much smaller crystal sizes than X-ray diffraction, this opens up the possibility to investigate the local structure of many technologically relevant materials for which no crystals large enough for single-crystal X-ray diffraction are available.

**Acknowledgements**: The authors would like to thank Dr. Nikolaj Roth for fruitful discussions and Dr. Lukas Palatinus for providing an option to apply symmetry averaging in the three-dimensional reciprocal lattice in *PETS2*.

**Author contributions**: Romy Poppe: Data collection; data processing; visualisation; writing – original draft; writing – review and editing. Joke Hadermann: writing – review and editing.

**References:**

Boysen, H. & Adlhart, W. (1987). *J. Appl. Crystallogr.* **20**, 200–209.

Brázda, P., Palatinus, L., Drahokoupil, J., Knížek, K. & Buršík, J. (2016). *J. Phys. Chem. Solids*. **96**, 10–16.

Fujii, Y., Miura, H., Suzuki, N., Shoji, T. & Nakayama, N. (2007). *Solid State Ionics*. **178**, 849–857.

Gemmi, M., Mugnaioli, E., Gorelik, T. E., Kolb, U., Palatinus, L., Boullay, P., Hovmöller, S. & Abrahams, J. P. (2019). *ACS Cent. Sci.* **5**, 1315–1329.

Gimenez, E. N., Ballabriga, R., Campbell, M., Horswell, I., Llopart, X., Marchal, J., Sawhney, K. J.

S., Tartoni, N. & Turecek, D. (2011). *IEEE Trans. Nucl. Sci.* **58**, 323–332.

Goodwin, A. L., Withers, R. L. & Nguyen, H. B. (2007). *J. Phys. Condens. Matter.* **19**, 335216.

Gorelik, T. E., Bekő, S. L., Teteruk, J., Heyse, W. & Schmidt, M. U. (2023). *Acta Crystallogr. Sect. B Struct. Sci. Cryst. Eng. Mater.* **79**, 122–137.

Jakůbek, J. (2009). *J. Instrum.* **4**,.

Kolb, U., Gorelik, T., Kübel, C., Otten, M. T. & Hubert, D. (2007). *Ultramicroscopy.* **107**, 507–513.

Kolb, U., Gorelik, T. & Otten, M. T. (2008). *Ultramicroscopy.* **108**, 763–772.

Kraft, P., Bergamaschi, A., Broennimann, C., Dinapoli, R., Eikenberry, E. F., Henrich, B., Johnson, I., Mozzanica, A., Schlepütz, C. M., Willmott, P. R. & Schmitt, B. (2009). *J. Synchrotron Radiat.* **16**, 368–375.

Krysiak, Y., Barton, B., Marler, B., Neder, R. B. & Kolb, U. (2018). *Acta Crystallogr. Sect. A Found. Adv.* **74**, 93–101.

Krysiak, Y., Marler, B., Barton, B., Plana-Ruiz, S., Gies, H., Neder, R. B. & Kolba, U. (2020). *IUCrJ.* **7**, 522–534.

Neagu, A. & Tai, C. W. (2017). *Sci. Rep.* **7**, 1–12.

Palatinus, L., Brázda, P., Boullay, P., Perez, O., Klementová, M., Petit, S., Eigner, V., Zaarour, M. & Mintova, S. (2017). *Science (80-. ).* **355**, 166–169.

Palatinus, L., Brázda, P., Jelínek, M., Hrdá, J., Steciuk, G. & Klementová, M. (2019). *Acta Crystallogr. Sect. B Struct. Sci. Cryst. Eng. Mater.* **75**, 512–522.

Pangaud, P., Basolo, S., Boudet, N., Berar, J. F., Chantepie, B., Clemens, J. C., Delpierre, P., Dinkespiler, B., Medjoubi, K., Hustache, S., Menouni, M. & Morel, C. (2008). *Nucl. Instruments Methods Phys. Res. Sect. A Accel. Spectrometers, Detect. Assoc. Equip.* **591**, 159–162.

Poppe, R., Vandemeulebroucke, D., Neder, R. B. & Hadermann, J. (2022). *IUCrJ.* **9**, 695–704.

Proffen, T. & Neder, R. B. (1997). *J. Appl. Crystallogr.* **30**, 171–175.

Roth, N., Beyer, J., Fischer, K. F. F., Xia, K., Zhu, T. & Iversen, B. B. (2021). *IUCrJ.* **8**, 695–702.

Schaub, P., Weber, T. & Steurer, W. (2007). *Philos. Mag.* **87**, 2781–2787.

Schmidt, E. M., Klar, P. B., Krysiak, Y., Svora, P., Goodwin, A. L. & Palatinus, L. (2023). 1–8.

Simonov, A., Weber, T. & Steurer, W. (2014). *J. Appl. Crystallogr.* **47**, 1146–1152.

Vincent, R. & Midgley, P. a. (1994). *Ultramicroscopy.* **53**, 271–282.

Weber, T. & Simonov, A. (2012). *Zeitschrift Fur Krist.* **227**, 238–247.

Welberry, T. R. & Goossens, D. J. (2016). *IUCrJ.* **3**, 309–318.

Welberry, T. R. & Weber, T. (2016). *Crystallogr. Rev.* **22**, 2–78.

Withers, R. L., Welberry, T. R., Brink, F. J. & Norén, L. (2003). *J. Solid State Chem.* **170**, 211–220.

Withers, R. L., Welberry, T. R., Larsson, A. K., Liu, Y., Norén, L., Rundlöf, H. & Brink, F. J. (2004). *J. Solid State Chem.* **177**, 231–244.

Xia, K., Nan, P., Tan, S., Wang, Y., Ge, B., Zhang, W., Anand, S., Zhao, X., Snyder, G. J. & Zhu, T. (2019). *Energy Environ. Sci.* **12**, 1568–1574.

Yu, J., Fu, C., Liu, Y., Xia, K., Aydemir, U., Chasapis, T. C., Snyder, G. J., Zhao, X. & Zhu, T. (2018). *Adv. Energy Mater.* **8**, 1–8.

Zeier, W. G., Anand, S., Huang, L., He, R., Zhang, H., Ren, Z., Wolverton, C. & Snyder, G. J. (2017). *Chem. Mater.* **29**, 1210–1217.

Zhao, H., Krysiak, Y., Hoffmann, K., Barton, B., Molina-Luna, L., Neder, R. B., Kleebe, H. J., Gesing, T. M., Schneider, H., Fischer, R. X. & Kolb, U. (2017). *J. Solid State Chem.* **249**, 114–123.